\documentclass[twocolumn]{openjournal}

\usepackage[T1]{fontenc}
\usepackage{ae,aecompl}
\usepackage[dvipsnames]{xcolor}
\usepackage{graphicx}	% Including figures
\usepackage{amsmath}	% Advanced maths commands
\usepackage{amssymb}	% Extra maths symbols

\usepackage{natbib}

\usepackage{hyperref}
\hypersetup{
	colorlinks=true,
	urlcolor=blue,    % color of external links
	linkcolor=black,  % color of toc, list of figs etc.
	citecolor=blue,   % color of links to bibliography
}

\defcitealias{Pittordis_2018}{PS18}
\defcitealias{Pittordis_2019}{PS19}
\defcitealias{Badry_2021}{ERH}
\defcitealias{Banik_2018_Centauri}{BZ18}

\hypersetup{citecolor=blue, % color for \cite{...} links
            linkcolor=red, % color for \ref{...} links
            menucolor=blue, % color for Acrobat menu buttons
            urlcolor=blue}  % color for \url{...} links

\newcommand{\km}{{~\rm km}}
\newcommand{\s}{{~\rm s}}

\newcommand{\AU}{{~\rm AU}}

\begin{document}

\title{The runaway velocity of the white dwarf companion in the double detonation scenario of supernovae}

%\correspondingauthor{Ealeal Bear, Noam Soker}
%\email{soker@physics.technion.ac.il}

%\author[0009-0001-4877-1125]{Jessica Braudo}
\author{Jessica Braudo, Noam Soker}
\affiliation{Department of Physics, Technion, Haifa, 3200003, Israel; jessicab@campus.technion.ac.il; soker@physics.technion.ac.il}

%\author[0000-0003-0375-8987]{Noam Soker}
%\author{Noam Soker}
%\affiliation{Department of Physics, Technion, Haifa, 3200003, Israel; jessicab@campus.technion.ac.il; soker@physics.technion.ac.il}

\begin{abstract}
We consider the finite velocity of the ejecta of a type Ia supernova (SN Ia) in the double detonation (DDet) scenario with a white dwarf (WD) mass-donor companion, and find that the runaway velocity of the surviving (mass donor) WD is lower than its pre-explosion orbital velocity by about $8-11 \%$.  This implies that the fastest runaway WDs in the Galaxy, if come from the DDet scenario, require even more massive WDs than what a simple calculation that neglects the finite ejecta velocity gives. This extreme set of initial conditions makes such binaries less common. We also tentatively find that the inner ejecta deviates from spherical symmetry,  but not to the degree that we can use observations to make any claim. Our findings support the claim that the DDet scenario leads mostly to peculiar SNe Ia but not to normal SNe Ia.  
\end{abstract}


% ==========================================================
\section{INTRODUCTION}
\label{sec:intro}
% ==========================================================

A number of theoretical scenarios of type Ia supernovae (SNe Ia) and peculiar SNe Ia exist (for recent reviews, see., e.g., \citealt{Hoeflich2017, LivioMazzali2018, Soker2018Rev, Soker2019Rev, Wang2018,  Jhaetal2019NatAs, RuizLapuente2019, Ruiter2020, Liuetal2023Rev}). 
\cite{Soker2019Rev} compared some scenarios to each other and by their fitting to observations. In Table \ref{Tab:Table1} we present four rows of a much larger table from that review. We present only the basic properties of the scenarios and their contribution to normal and peculiar SNe Ia as \cite{Soker2019Rev} crudely estimated. 
 The properties (second row) are $N_{\rm exp}$, which is the number of stars in the system at the time of explosion, $N_{\rm sur}$, which is the number of surviving stars after the explosion, and the mass of the exploding white dwarf (WD), $M_{\rm Ch}$ for near Chandrasekhar mass WD and sub-$M_{\rm Ch}$ for a sub-Chandrasekhar exploding WD. 
% TTTTTTTTTTTTTTTTTTTTTTTTTTTTTTTTTTTTTTTTTTTTTTTTTTTTTT
\begin{table*}
%\tiny
\scriptsize
%\footnotesize
\begin{center}
  \caption{SN Ia scenarios}
    \begin{tabular}{| p{2.5cm} | p{2.1cm}| p{2.1cm}| p{2.1cm}| p{2.1cm} | p{2.1cm} | p{2.1cm} |}
\hline  % ----------------------------
\textbf{{Scenario$^{[{\rm 1}]}$}}  & {Core Degenerate \newline (CD)}    & {Double Degenerate} (DD) & {Double Degenerate} (DD-MED)& {Double Detonation} (DDet) & {Single Degenerate} (SD-MED) & {WD-WD collision} (WWC)\\
\hline  % ----------------------------
 {$\mathbf{[N_{\rm exp}, N_{\rm sur}, M, Ej]}$$^{[{\rm 2}]}$}
  & $[1,0,M_{\rm Ch},{\rm S}]$ 
  & $[2,0,$sub-$M_{\rm Ch},{\rm N}]$
  & $[1,0,M_{\rm Ch}, {\rm S}]$ 
  & $[2,1,$sub-$M_{\rm Ch},{\rm N}]$
  & $[2,1,M_{\rm Ch},{\rm S}]$  
  & $[2,0,$sub-$M_{\rm Ch},{\rm N}]$ \\
\hline  % ----------------------------
{Contribution to normal SNe Ia$^{[{\rm 5}]}$} 
 & $\approx 20-50 \%$ 
 & $\approx 20-40 \%$ 
 & $\approx 20-40 \%$
 & $\approx 0-10 \%$ 
 & $\approx 0-10 \%$ 
 & $\ll 1 \%$\\
\hline  % ----------------------------
{Contribution to peculiar SNe Ia}
 & $\approx 0-10 \%$
 & $\approx 30-70 \%$
 & $\approx 0-10 \%$
 & $\approx 10-30 \%$ 
 & $\approx 20-50 \%$ by the SD scenario without MED  
 & $\approx 1 \%$ \\
\hline  % ----------------------------

%strong constrains on the properties of a possible companion
     \end{tabular}
  \label{Tab:Table1}\\
\end{center}
\begin{flushleft}
\small SN Ia scenarios and the crude estimate by \cite{Soker2019Rev} of their contributions to normal and peculiar SNe Ia. \\
Notes: \small [1] Scenarios for SN Ia by alphabetical order. MED: Merger to explosion delay time. It implies that the scenario has a delay time from merger or mass transfer to explosion. MED is an integral part of the CD scenario.   
 \\
 \small [2] $N_{\rm exp}$ is the number of stars in the system at the time of the explosion; $N_{\rm sur}=0$ if no companion survives the explosion while $N_{\rm sur}=1$ if a companion survives the explosion; $M_{\rm Ch}$ indicates a (near) Chandrasekhar-mass explosion while sub-$M_{\rm Ch}$ indicates sub-Chandrasekhar mass explosion; Ej stands for the morphology of the ejecta, where S and N indicate whether the scenario might lead to spherical explosion or cannot, respectively.   
\end{flushleft}
\end{table*}
% TTTTTTTTTTTTTTTTTTTTTTTTTTTTTTTTTTTTTTTTTTTTTTTTTTTTTTTTTTTTTTTT

Since the community is far from any consensus on the dominate scenarios for normal and peculiar SNe Ia, as evident by the diversity of recent studies on the different scenarios (e.g., some recent papers from 2022 on, \citealt{Ablimit2022, Acharovaetal2022, AlanBilir2022, Barkhudaryan2022, Chanlaridisetal2022, Chuetal2022, CuiLi2022, Dimitriadisetal2022, Ferrandetal2022,  Kooletal2022, Kosakowskietal2022, Kwoketal2022, Lachetal2022, Liuetal2022, LivnehKaatz2022, Mazzalietal2022, Pakmoretal2022, Patraetal2022, Piersantietal2022, RauPan2022, RuizLapuenteetal2022, Sanoetal2022, SharonKushnir2022, Shingles2022, Tiwarietal2022, DerKacyetal2023a, DerKacyetal2023b, Igoshevetal2023, IwataMaeda2023, Kobashietal2023,  LaversveilerGoncalves2023, LiuJetal2023, MoranFraileetal2023, SwarubaRajamuthukumaretal2023, WangLetal2023, WangQetal2023, Chakrabortyetal2024, Ritteretal2023}), it is mandatory to present all these binary scenarios. 
\begin{enumerate}
\item The \textit{core-degenerate (CD) scenario} leaves no surviving companion and predicts a large-scale spherical explosion; small-scale non-homogeneous structures are expected. It involves a common envelope evolution (CEE) ending with the merger of the CO (or HeCO) core of a massive asymptotic giant branch (AGB) star with a CO (or HeCO) WD. The $M_{\rm Ch}$-WD remnant of the merger explodes later (e.g., \citealt{KashiSoker2011, Ilkov2013, AznarSiguanetal2015}). 
This scenario includes a merger to explosion delay (MED) time. 
\item The \textit{double degenerate (DD) scenario} (e.g., \citealt{Webbink1984, Iben1984}) leaves no companion and predicts non-spherical explosion because explosion takes place as the two WDs merge, i.e., no MED time. There are different channels of the DD scenario (e.g., \citealt{Pakmoretal2011, Liuetal2016, Ablimitetal2016}), including the ignition of helium first (e.g., \citealt{YungelsonKuranov2017, Zenatietal2019, Peretsetal2019}). Because it predicts non-spherical ejecta, this scenario fits better peculiar SNe Ia than normal SNe Ia, like calcium-rich SNe (e.g., \citealt{Zenatietal2023}). In recent studies of the peculiar SN 2022pul \cite{Kwoketal2023} and \cite{Siebertetaletal2023b} argue that the violent merger channel of the DD scenario can account for this ``super-Chandrasekhar'' SN Ia (for other related recent studies see, e.g.,  \citealt{AxenNugent2023, Maedaetal2023, Siebertetal2023a, Srivastavetal2023}).  
\item The \textit{DD-MED scenario} starts as the DD scenario but there is a MED time, i.e., the explosion takes place only sometime after the merger (e.g., \citealt{LorenAguilar2009, vanKerkwijk2010, Pakmor2013, Levanonetal2015, LevanonSoker2019, Neopaneetal2022}). This scenario predicts a spherical explosion which is compatible with the low polarization of some SNe Ia and with the large-scale spherical structure of some SN Ia remnants. 
\item In the \textit{double-detonation (DDet) scenario} (e.g., \citealt{Woosley1994, Livne1995, Shenetal2018b}) the explosion of a CO WD is triggered by the thermonuclear burning of a helium layer (e.g., \citealt{Zingaleetal2023} for a recent paper) that the CO WD accretes from a companion. It has several channels, (e.g., the core merger detonation scenario; \citealt{Ablimit2021}). In most channels the companion survives, but not in all, e.g., in the triple detonation scenario (e.g., \citealt{Papishetal2015}). The channel with a surviving WD mass donor is the subject of the present study.  
\item In the \textit{single degenerate (SD) scenario} the $M_{\rm Ch}$ exploding CO WD accretes hydrogen-rich gas from a non-degenerate mass-donor (e.g., \citealt{Whelan1973, HanPodsiadlowski2004, Orio2006, Wangetal2009, MengPodsiadlowski2018}). Several types of donors are possible even inside a common envelope (the common-envelope wind model; e.g., \citealt{Cuietal2022}). As well, the explosion might take place as soon as the CO WD reaches close to the Chandrasekhar mass limit or it might occur with a long time delay after the CO WD losses angular momentum (e.g., \citealt{Piersantietal2003, DiStefanoetal2011, Justham2011}). The latter is the SD-MED 
scenario. In the SD without a MED time, the companion will block some of the ejecta, e.g., a giant star at $\simeq 1 \AU$, hence leading to a non-spherical ejecta (e.g., \citealt{Liuetal2013}). However, in the SD-MED scenario, the donor is a giant star that by the time of the explosion is a WD at an orbital separation of $\simeq 1 \AU$. Such a WD has a very small influence on the morphology of the ejecta. The explosion can be spherical on a large scale.   
\item  In the \textit{WD-WD collision (WWC) scenario} a collision of two CO WDs ignites the explosion (e.g., \citealt{Raskinetal2009, Rosswogetal2009, Kushniretal2013, AznarSiguanetal2014}). It leaves no companion and predicts a highly non-spherical explosion. The WWC scenario accounts for   $<1 \%$ of normal SNe Ia (e.g., \citealt{Toonenetal2018, HallakounMaoz2019, HamersThompson2019, GrishinPerets2022}).  It might account for some peculiar SNe Ia.   
\end{enumerate}

\cite{Soker2019Rev} estimated that the DDet scenario might account for a large fraction of peculiar SNe Ia, but for only a very small fraction of normal SNe Ia. The reasons are that the DDet scenario results in highly-non-spherical ejecta, contrary to many normal SN Ia remnants and that no companions were found in SN Ia remnants (e.g., \citealt{LiuDetal2019, Shieldsetal2022, Shieldsetal2023ApJ}). \cite{Liuetal2023} and \cite{PadillaGonzalezetal2023} argue that the DDet scenario explains the peculiar SN Ia SN 2022joj. \cite{KYadavallietal2023} suggest the DDet scenario for the peculiar SN 2022oqm. Moreover, \cite{Royetal2022} show in their simulations that the detonation of the helium layer does not always detonate the CO WD. They conclude that this suggests that the DDet scenario with a WD companion might be limited to the most massive CO primary.

In some channels of the DDet scenario, e.g., the D6 (for ``dynamically driven double-degenerate double-detonation''; \citealt{Shenetal2018a})
channel (e.g., \citealt{Finketal2010, Guillochonetal2010}), the helium-donor companion is a WD. After the explosion the surviving WD has a high velocity relative to the center of mass of the pre-explosion binary, hundreds to thousands of $\km \s^{-1}$. The existence of hyper-runaway/hypervelocity WDs is a prediction of the D6 model. \cite{Igoshevetal2023}  analyze Gaia data and conclude that the numbers and properties of hypervelocity WDs with velocities of $ >1000 \km \s^{-1}$ might at most account for a small fraction of SNe Ia. They argue that this finding disfavors a significant contribution of the D6 channel to SNe Ia. Nonetheless, as stated, the DDet scenario can account for some peculiar SNe Ia. 
\cite{Shenetal2018a} discover three hypervelocity WDs and argue that each is a remnant of the helium donor in the DDet scenario (also \citealt{Baueretal2021}). 
\cite{ElBadryetal2023} analyze Gaia data and find several more hypervelocity WDs. The fastest WDs that have inferred birth velocities of $\simeq 2200 - 2500 \km \s^{-1}$ imply massive donor and exploding WDs. They conclude that their statistical analysis is consistent with the possibility that a large fraction and even all SNe Ia are produced via the D6 channel.  

As we indicated above, there are other arguments against the D6 scenario as a significant contributor to the population of normal SNe Ia, but might be a major contributor to peculiar SNe Ia. For that, there is an interest in better understanding the properties of the surviving helium donor. Some studies take the birth velocity of the ejected helium donor to be the pre-explosion Keplerian velocity. In this study, we examine the value of the birth velocity and the implication on the departure of the inner ejecta from spherical symmetry. We describe our calculation scheme in section \ref{sec:Numerical}. We present the results concerning the birth hypervelocity and the structure of the ejecta in sections \ref{sec:Ruaway} and \ref{sec:Nonspherical}, respectively. We summarize our study in section \ref{sec:Summary}.  

% =====================================================
\section{The numerical procedure}
\label{sec:Numerical}
% =====================================================
We wrote a MATLAB code to evolve binary WD systems for 120 seconds starting with the explosion of one WD, the mass-accretor WD. At explosion, the mass-accretor WD mass is either $1.07M_{\odot}$ or $1.2M_{\odot}$, and the mass-donor WD mass is either $0.9M_{\odot}$ or $1.05M_{\odot}$. In the presently studied DDet scenario, the accretor is the exploding WD and the donor is the surviving WD. We take into account the structure of the ejecta at its terminal velocity. The ejecta model is taken from \cite{Gronowetal2021} and the Heidelberg Supernova Model Archive (\textsc{HESMA}\footnote{https://hesma.h-its.org}, e.g., \citealt{Kromeretal2017}). The ejecta is composed of 85 numerical mass shells expanding with terminal velocities in the range of $670 \km \s^{-1} - 33,000 \km \s^{-1}$.

The Roche lobe overflow (RLOF) by the donor with a radius $R_{\rm sur}$, takes place at an orbital separation of (\citealt{Eggleton1983})
\begin{equation}
    a_0=\frac{R_{\rm sur}}{0.49}
    \left[ 0.6 + q^{-2/3} \ln (1+q^{1/3})   \right],  
\label{eq:RLOF}
\end{equation}
where $q=M_{\rm sur}/M_{\rm exp}$ is the mass radio of the surviving WD (the mass-donor WD) to the exploding WD (the mass-accretor WD) $M_{\rm exp}$. 
The orbital velocity of the surviving WD around the center of mass just before the explosion is 
\begin{equation}
    v_{\rm orb,s} = \sqrt{ 
    \frac{G M_{\rm exp}}{(1+q) a_0}  
    }    .
\label{eq:VorbS}
\end{equation}

Given the densities of the mass shells, we calculate their masses and include these in the calculation of the velocity of the surviving WD as a function of time in the following way. We assume that each shell maintains its spherical structure. We solve for the orbital motion of the surviving WD around the ejecta shells that have not crossed it yet. Namely, each expanding ejecta shell that engulfs the surviving WD is removed from the calculation of the surviving WD velocity. Only numerical mass shells that did not cross the surviving WD continue to influence the motion of the surviving WD. 
Because we do not include hydrodynamical effects, each numerical mass shell that engulfs the surviving WD is not considered anymore, even if later the WD exits that shell due to its high velocity.

We take each time step to be $\Delta t = 0.001 \s$. We checked also $\Delta t = 0.004 \s$ and $\Delta t=0.00025 \s$ and found changes of only 0.004\% in the final WD velocity.

A comment is in place here on the usage of equation (\ref{eq:RLOF}) for the orbital separation at RLOF. Simulations show that mass transfer in binary WD systems starts at a somewhat larger (by several percent) orbital separation than what equation (\ref{eq:RLOF}) gives (e.g., \citealt{Danetal2011, Pakmoretal2012}). If we use a somewhat large separation at explosion, then to obtain a given runaway velocity the masses of the two stars should be larger even. The larger separation would imply also that the effect we study here is large because the longer time the ejecta takes to expand beyond the surviving WD. Both these effects will constrain the WDs to be more massive than what we find here, strengthening our final conclusion.

We also consider the momentum that the ejecta shells impart to the WD donor (kick; e.g., \citealt{ElBadryetal2023}). The change in the donor's velocity due to the momentum that shell number $i$ imparts to the surviving WD is given by  
\begin{equation}
    v_{\rm kick,i} = \eta \frac{\pi R^2_{\rm sur}}{4 \pi [a(t)]^2} 
    \frac{p_{\rm ej,i}}{M_{\rm sur}},   
\label{eq:vkick}
\end{equation}
where $\eta = 0.5 $ is a momentum transfer efficiency factor, as calculated by, e.g.,  \cite{ElBadryetal2023}, $p_{\rm ej,i}$ is the radial momentum of ejecta shell $i$, and $a(t)$ is the distance of the surviving WD from the explosion site at the moment ejecta shell $i$ of mass $M_{\rm ej, i}$ hits it. 

When the ejecta hits the surviving WD it is not yet at its terminal velocity because at early times the ejecta is still hot, and not all thermal energy has been channeled to kinetic energy.  To account for this we also consider the same ejecta model but with the velocity of each of the 85 numerical mass shells taken to be a fraction of $\beta=0.85, 0.9$ or $0.95$ of its terminal velocity. The terminal velocity case corresponds to $\beta=1$. 

% =====================================================
\section{The runaway velocity}
\label{sec:Ruaway}
\begin{figure*}[]
	\centering
%	\hspace*{-2cm} 
	% This cut edges: [trim=left bottom right top, clip]{file}
%	\hspace{1cm}
\includegraphics[trim=0.0cm 0.0cm 0.0cm 0.0cm ,clip, scale=0.75]{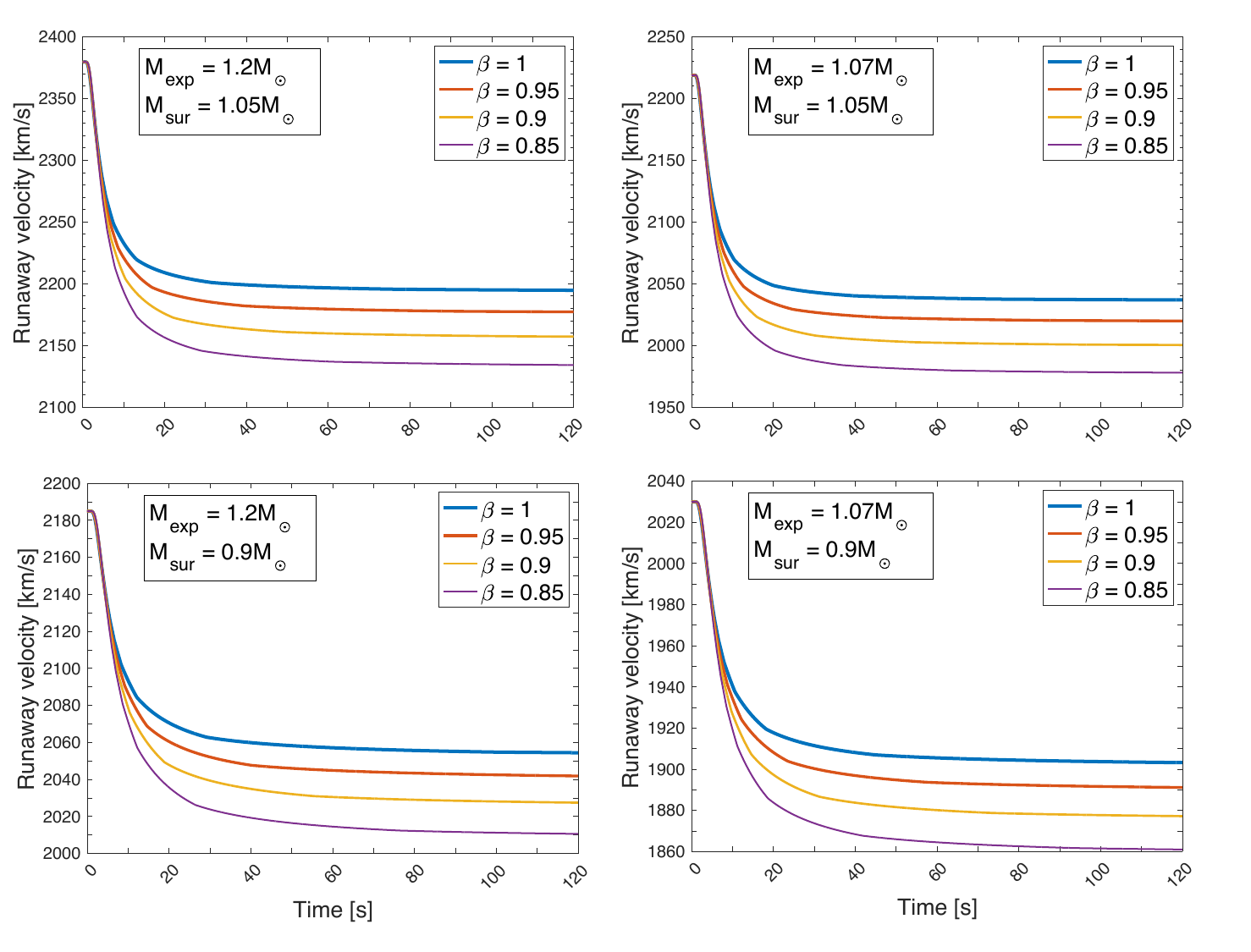} 
\caption{The runaway velocity of the surviving WD as a function of time after SN Ia explosion in the DDet scenario. The inset in each panel shows the masses of the exploding (mass-accretor) WD $M_{\rm exp}$ and of the surviving (mass-donor) WD $M_{\rm sur}$. The different lines depict cases with different ratios of the ejecta early velocity to its terminal velocity $\beta$. 
}
\label{Fig:Velocities}
\end{figure*}

% =====================================================
We numerically calculated the runaway velocities in 16 different cases for different values of $\beta$ (ratio of ejecta velocity to its terminal velocity) and different donor-to-accretor mass ratios $q$ (section \ref{sec:Numerical}). In Fig. \ref{Fig:Velocities} we present the mass-donor WD (the surviving WD) velocity, i.e. its runaway velocity, as a function of time during the 120 seconds of the simulation. At $t=120 \s$ the runaway WD about reaches its terminal runaway velocity. 
Each panel presents the results for different masses of the exploding and/or surviving WD (insets), and each panel presents the four values of $\beta$ as indicated for the different-colored lines.  
% FFFFFFFFFFFFFFFFFFFFFFFFFFFFFFFFFFFFFFFFFFF  
%FFFFFFFFFFFFFFFFFFFFFFFFFFFFFFFFFFFFFFFFFF

The initial velocity in each panel, i.e., at $t=0$, is the pre-explosion Keplerian velocity of the surviving WD. We learn from the different lines that because the ejecta requires time to engulf (cross) the surviving WD, the surviving WD slows down due to its gravitational interaction with the ejecta that has not crossed (engulfed) it yet. For all masses, the final runaway velocity decreases with decreasing $\beta$. Namely, a slower expansion velocity, because not all the thermal energy of the ejecta was channeled to kinetic energy yet, implies that the mass shells of the ejecta take more time to expand and engulf the donor WD. The gravitational interaction lasts longer and therefore the shells' gravitational force slows down the runaway WD more. For the cases we simulated we find that typically the final runaway velocity is $\simeq 8-11\%$ slower than the initial orbital velocity.

The initial orbital velocities and the final runaway velocities are higher in cases with larger masses. Only such cases can produce very fast observed runaway velocities. For example, \cite{ElBadryetal2023} explain the observed runway velocity $2372 \km \s^{-1}$ of J0927-6335  with a binary system of  $M_{\rm exp} \simeq 1.2M_{\odot}$ and $M_{\rm sur} \simeq 1.05M_{\odot}$. This is the pre-explosion orbital velocity of the surviving WD. Our results (upper left panel) show that the final runaway velocity for this case, even for $\beta=1$, is only $v_{\rm sur}=2200 \km \s^{-1}$. In this case, the kick velocity, i.e., the momentum that the ejecta imparts directly to the surviving WD, increases the final runaway velocity only by 0.009\% and it is therefore negligible.

The main conclusion from the discussion above is that even more extreme (more massive) WDs are required to explain the fastest runaway WDs. 
 \cite{Royetal2022} have reached a similar conclusion based on their study of the denotation of the CO WD by the detonation of the helium layer.
The requirement for very massive CO WDs does not rule out the DDet scenario for these systems but makes it less common. Considering that \cite{Igoshevetal2023} find that the rate of observed hypervelocity WDs with velocities of $ >1000 \km \s^{-1}$ implies that SN Ia scenarios that predict such hypervelocity WDs account for at most a small fraction of SNe Ia, our results implies even a lower fraction that what \cite{Igoshevetal2023} claim for. We here strengthen their claim.
We argue that this is compatible with the fact that the DDet scenario accounts mostly (or even only) for peculiar SNe Ia (but not for all peculiar SNe Ia), but does not account for normal SNe Ia (see section \ref{sec:intro} and Table \ref{Tab:Table1}).

% ==========================================
\section{A non-spherical inner region}
\label{sec:Nonspherical}
% ==========================================

\begin{figure*}[]
	\centering
%	\hspace*{-2cm} 
	% This cut edges: [trim=left bottom right top, clip]{file}
%	\hspace{1cm}
\includegraphics[trim=3.2cm 7.2cm 0.0cm 7.8cm ,clip, scale=1.2]{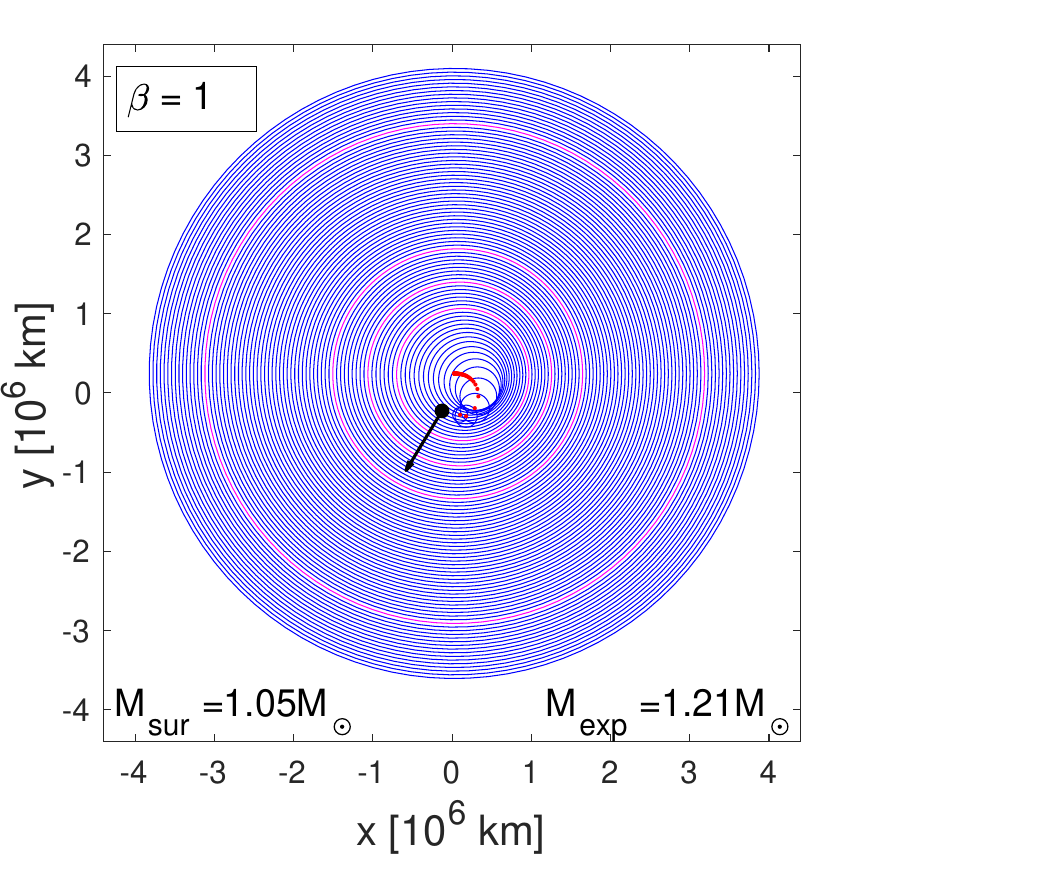} 
\caption{The location of 82 spherical numerical mass shells on the equatorial plane at $t=120 \s$ after the explosion for the case with $\beta=1$, $M_{\rm exp}=1.21M_{\odot}$ and $M_{\rm sur}=1.05M_{\odot}$. The black dot shows the location of the mass-donor (surviving) WD at $t=120 \s$ and the black arrow indicates the WD velocity direction. Magenta circles show ejecta mass coordinates of $0.3M_{\odot}$, $0.6M_{\odot}$, $0.9M_{\odot}$ and $1.2M_{\odot}$. The total ejecta mass is $M_{\rm exp}=1.21M_{\odot}$. Each of the 82 red points represents the center of one mass shell of the ejecta at $t=120 \s$. The upper most red dot is the center of the outer most shell. These 82 points and the shells themselves show that the inner parts largely deviate from spherical symmetry. The inner three numerical mass shells did not engulf (cross) the surviving WD, thus we present only 82 numerical mass shells out of the 85 shells that compose the ejecta.
}
\label{Fig:Spherres}
\end{figure*}

We recall that our simulated ejecta is composed of 85 numerical mass shells. We followed the expanding numerical mass shells for the 120 seconds of the simulation. In Fig. \ref{Fig:Spherres} we preset the locations of the outer 82 numerical mass shells on the orbital plane for the case in which $\beta = 1$, $M_{\rm exp} = 1.2M_\odot$ and $M_{\rm sur} = 1.05M_{\odot}$. We also mark the location and velocity direction of the surviving WD at $t=120 \s$. The three inner numerical shells that did not engulf the WD during the simulation are not shown. In the different 16 simulated cases between one to three inner shells did not engulf the WD by $t=120 \s$.
The coordinate $(x,y,z)=(0,0,0)$ is set at $t=0$, which is the explosion time, to be the center of mass of the binary system. At $t=0$ the exploding WD is at $(x,y,z)=(-7057 \km,0,0)$ and the surviving WD is at $(x,y,z)=(8134 \km,0,0)$.
% FFFFFFFFFFFFFFFFFFFFFFFFFFFFFFFFFFFFFFFFFFF  
%FFFFFFFFFFFFFFFFFFFFFFFFFFFFFFFFFFFFFFFFFF

Note that some ejecta mass shells engulf the WD, but later the WD exists from these shells because of its fast runaway velocity. We can see in Figure \ref{Fig:Spherres} that the surviving WD (black dot) is indeed outside some shells. 

Outer numerical mass shells have velocities much higher than the initial orbital velocity of the exploding star. They are not influenced much by the gravity of the surviving WD and expand with high (almost) constant velocity throughout the simulation. Inner numerical mass shells are influenced more. During the expansion and before engulfing the surviving WD, these shells affect the surviving WD motion and, by Newton's third law, are affected by the surviving WD. Before the outermost numerical shell crosses the surviving WD the ejecta and the surviving WD continue their orbital motion. As more numerical mass shells engulf (cross) the surviving WD and expand beyond it, the ejecta that is affected by the surviving WD, and hence affect the motion of the surviving WD, decreases. These inner numerical mass shells that expand at slower velocities are deflected more by the surviving WD. We see in Fig. \ref{Fig:Spherres} that the inner numerical mass shells do not share the same center as the outer ones, with departure between the centers of numerical mass shells increasing to inner shells. Both the location of the numerical mass shells on the equatorial plane and the dots, which are the centers of these 82 numerical mass shells, show this. We find very similar deviations from symmetry in the other 16 simulated cases.  

The main conclusion from the above finding is that the departure from spherical symmetry of the inner parts of the ejecta when the two WDs are massive is non-negligible. The departure is not large enough and observations are not sufficiently detailed to allow for a comparison between them. Nonetheless, the departure might suggest a peculiar SN Ia, as most SNe Ia show generally spherical ejecta. More accurate hydrodynamical simulations are required to obtain the exact morphology of the ejecta.

% ==========================================
\section{Summary}
\label{sec:Summary}
% ==========================================

We studied some effects of the finite velocity of the SN Ia ejecta on the final velocity of the surviving (mass-donor, mainly helium-donor) WD in the DDet scenario with a WD companion. We considered that the expansion velocity of the ejecta increases with mass coordinate. Namely, before the inner ejecta mass layers cross (engulf) the surviving WD, the WD continues its orbital motion around the inner ejecta. 

One outcome, as we show in Figure \ref{Fig:Velocities}, is that the final runaway velocity of the surviving WD for massive WD binary systems is slower than its pre-explosion orbital velocity by about $8-11 \%$. This implies that the fastest runaway WDs in the Galaxy (e.g., \citealt{ElBadryetal2023}) require even more massive WDs than what a simple calculation that neglects the finite ejecta velocity gives. This makes the initial conditions less common. \cite{Igoshevetal2023} found that SN Ia scenarios that leave hypervelocity WDs can at most contribute a small fraction of normal SNe Ia. Our finding further reduces this rate and therefore strengthens the conclusion of \cite{Igoshevetal2023}.

Our second finding is that the inner ejecta largely deviates from spherical symmetry. This finding, however, requires simulations that include hydrodynamical effects to confirm or reject our finding. 

We claim that our study supports the notion (section \ref{sec:intro} and Table \ref{Tab:Table1}) that the DDet scenario leads to peculiar SNe Ia but not to normal SNe Ia. 

We can take our results beyond just supporting the claim that the DDet scenario results in peculiar SNe Ia. Our conclusion also adds a small support to the idea that normal SNe Ia are due to the explosion of a lonely WD (for a review see \citealt{Soker2024Rev}). Namely, at explosion, there is only one star, the exploding WD that is the merger product of a core with a WD in the CD scenario, or of two WDs in the DD-MED scenario (see Table \ref{Tab:Table1}). This lonely-WD explosion explains the tight relation between peak luminosity and luminosity decline rate of normal SNe Ia. In this picture, the SNe Ia scenarios that have two stars at explosion (all scenarios besides the CD and the DD-MED) lead to peculiar SNe Ia. They are much richer in their variety of companions' properties, explaining the large area of peculiar SNe Ia in the plane of peak luminosity versus decline rate.

% ==========================================
\section*{Acknowledgments}
% ==========================================
We thank an anonymous referee for good comments. This research was supported by a grant from the Israel Science Foundation (769/20).

%%%%%%%%%%%%%%%%%%%%%%%%%%%
%\section*{Data availability}
%The data underlying this article will be shared on reasonable request to the corresponding author.  
%%%%%%%%%%%%%%%%%%%%%%%%%%%

% %%%%%%%%%%%%  References %%%%%%%%%%%%%%%%%%%%%

\end{document}